\documentclass[runningheads]{llncs}
\usepackage[scr=rsfs]{mathalpha}
\usepackage{graphicx}
\usepackage{amssymb}
\usepackage{amsmath}
\usepackage{algorithm}
\usepackage{algpseudocode}
\usepackage{orcidlink}
\usepackage{hyperref}
\hypersetup{
    hidelinks 
}

\usepackage{color}
\usepackage{subcaption}

\begin{document}
\title{User-Assisted Collaborative Distributed Inference for Efficient QoS-Aware Autoscaling}
\titlerunning{User-Assisted Collaborative Distributed Inference}

\author{Alfreds Lapkovskis\orcidlink{0009-0003-4424-949X} \and
Ali Beikmohammadi\orcidlink{0000-0003-4884-4600} \and \\
Sindri Magnússon\orcidlink{0000-0002-6617-8683} \and
Praveen Kumar Donta\orcidlink{0000-0002-8233-6071}}

\authorrunning{A. Lapkovskis et al.}

\institute{Department of Computer Systems and Sciences, Stockholm University \\ SE-106 91 Stockholm, Sweden \\ \email{\{alfreds.lapkovskis, beikmohammadi, sindri.magnusson, praveen\}@dsv.su.se}}

\maketitle              
\begin{abstract}
Growing demand for artificial intelligence (AI) inference services requires scalable infrastructure, yet centralized serving costs rise with demand. We propose a collaborative distributed inference system combining dedicated infrastructure with resources contributed by service users. Dedicated resources provide baseline capacity for maintaining quality of service (QoS), while volunteered resources absorb increasing demand without proportional growth in centralized infrastructure. To capture stochastic and dynamic interactions among users, resources, tasks, and policies, we develop a high-dimensional generative Markov model with structured temporal factorization. The model supports simulation and provides a foundation for task scheduling and QoS-aware resource allocation optimization. We evaluate the system across user populations, resource capacities, and centralized and distributed scheduling policies. Simulations show that distributed scheduling becomes increasingly advantageous as the user population grows, improving request completion and P99 latency while substantially reducing dedicated resource consumption. These results demonstrate the feasibility of user-assisted collaborative inference for infrastructure-efficient autoscaling.

\end{abstract}

\section{Introduction}

Artificial intelligence (AI) inference has become a core service in modern computing systems. According to the AI Index Report 2026, corporate AI adoption has reached 88\%, while global investments surged to \$581.69 billion in 2025 \cite{sajadieh2026aiindex}. As AI inference scales to meet this demand, cumulative inference costs increasingly outpace one-time training expenses, making operational efficiency a growing concern.
These trends highlight the importance of the computing paradigms underlying AI inference. Central to these paradigms is cloud computing,
the de facto standard for Internet applications \cite{mengistu2019survey}, including AI inference. 
Its key advantages include on-demand resource provisioning, rapid elasticity, standardized access interfaces, utilization monitoring, and pay-as-you-go pricing \cite{mell2011nist,ma2024research}.

However, centralized clouds are also subject to significant challenges. 
The costs associated with the setup and management of cloud infrastructure are prohibitively high \cite{mengitsu2018cucloud,mengistu2019survey}. Furthermore, expenses grow proportionally with demand, making cloud computing increasingly costly at scale. This scaling also carries a substantial environmental burden. Training Grok 4 alone emitted more CO\textsubscript{2} than an average car produces over its entire lifetime, and the annual inference cost of GPT-4o exceeds the drinking water demand of 12 million people \cite{sajadieh2026aiindex}. More broadly, cloud data centers consume vast amounts of energy and freshwater, with AI workloads emerging as a primary driver \cite{li2025making}, and these impacts are projected to worsen as AI demand continues to rise.

Edge and fog computing partially address these limitations. By distributing computation closer to end users and data sources, they reduce latency and bandwidth consumption, and decrease reliance on centralized cloud resources \cite{donta2023exploring}. The computing continuum extends this approach by integrating cloud, fog, and edge tiers into a unified, synergistic architecture~\cite{dustdar2023distributed}. Recent work has improved inference within these environments through heterogeneous on-device co-execution~\cite{11328908}, task-aware DNN partitioning in mobile edge computing~\cite{11313635}, and joint batching and transformer partitioning for distributed LLM inference~\cite{11328882}. Other studies address on-device inference techniques~\cite{11177548}, disaggregated inference deployment across heterogeneous infrastructure~\cite{11361142}, inference-data distribution for model monitoring~\cite{marzban2025inference}, and cloud inference autoscaling~\cite{10.1145/3727353.3727398}. These approaches improve inference within predefined deployment environments, but generally rely on dedicated infrastructure rather than resources that become available dynamically as users join the service. Moreover, computing-continuum systems still face architectural and economic barriers to widespread adoption~\cite{ding2022roadmap,meuser2024revisiting}.

Yet deploying dedicated edge and fog infrastructure is not the only path forward. A vast number of commodity devices
owned by individuals and institutions worldwide, remain largely underutilized \cite{che2011creditunion,mengitsu2018cucloud}. Harnessing these idle resources toward a shared computational objective is known as volunteer computing \cite{ma2024research,mengistu2019survey}, a paradigm that offers access to cheaper, distributed compute without requiring energy-hungry data center infrastructure. 
However, existing systems provide limited support for stable QoS guarantees due to resource heterogeneity and intermittent availability \cite{mengistu2019survey}.
Hence, this model has limited applicability to consumer-oriented AI inference applications. Yet the underlying concept
remains compelling in the context of AI inference, where operational costs are a growing concern.

\begin{figure} [!t]
    \centering
    \includegraphics[width=0.92\textwidth]{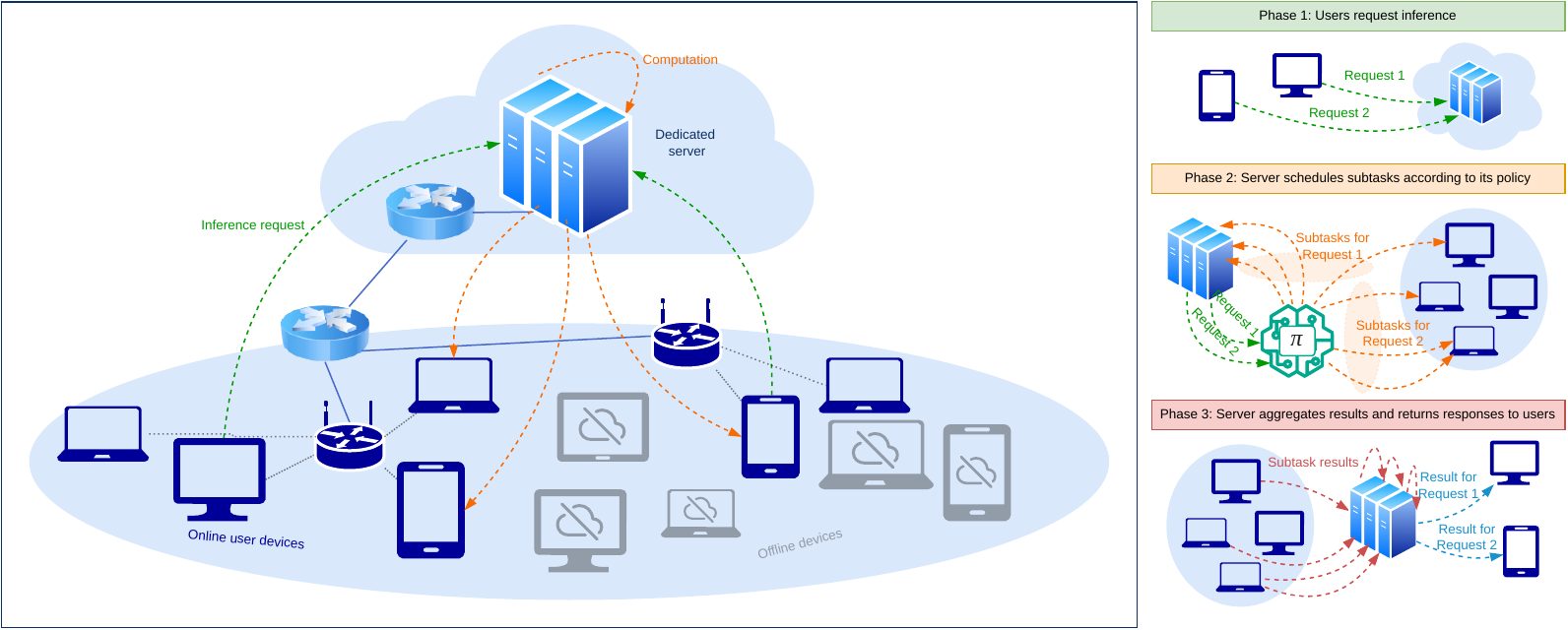}
    \caption{Overview of the proposed distributed inference system. The server receives requests, schedules subtasks across dedicated and volunteered resources according to the scheduling policy, and aggregates the results into responses.}
    \label{fig:system_overview}
\end{figure}

Therefore, we propose a collaborative distributed inference system that partitions inference computations between dedicated resources, such as centralized servers, and volunteered resources, such as users' devices (as shown in Fig.~\ref{fig:system_overview}). Unlike centralized cloud inference or approaches distributed exclusively across dedicated computing continuum infrastructure, our system does not require dedicated computational capacity to scale in proportion to demand. In contrast to conventional volunteer computing, however, it retains dedicated infrastructure to provide a reliable baseline capacity and support QoS guarantees despite the intermittent availability and heterogeneity of volunteered resources. Given an appropriate scheduler, the server can process all requests when few users are online; as the number of available users grows, the demand increases, but so does the pool of volunteered resources, allowing the system to progressively offload computation to user devices. Determining how to schedule tasks across these dynamic resources and how much dedicated capacity is necessary to satisfy QoS requirements requires a model that captures the complex interactions among users, resources, tasks, and policies. We therefore develop such a model to enable system analysis, scheduling optimization, and QoS-aware allocation of dedicated resources. Our contributions are threefold:

\begin{enumerate}
    \item We propose a novel collaborative distributed AI inference system that combines the dedicated and volunteered user resources for automatic scaling while maintaining QoS and reducing the need for centralized resources.

    \item We model the system as a generative Markov model whose temporal factorization captures sparse dependencies among structured state variables and their elements, enabling tractable simulation, inference, and policy learning.

    \item We evaluate and compare various scheduling policies by simulating the proposed system under diverse configurations, varying the number of users, dedicated, and volunteered resources.
\end{enumerate}

\section{Proposed Work}

We propose a collaborative distributed inference system that partitions inference computations between dedicated resources, such as centralized servers, and volunteered resources, such as users’ devices. 
Ensuring QoS in such a system requires not only provisioning additional dedicated resources but also optimizing computation scheduling strategies. In principle, QoS can be maintained either by increasing server capacity and executing more computations on dedicated infrastructure or by improving the scheduler to make more effective use of volunteered resources. However, volunteered resources are inherently heterogeneous, highly dynamic, and intermittently available. Scheduling decisions must therefore be adaptive and capable of anticipating future system states to prevent QoS degradation while efficiently utilizing the available resource pool. Optimizing such scheduling strategies directly on a live system is impractical: it would be costly, time-consuming, and require extensive trial-and-error exploration in a complex, dynamic environment. Consequently, a system model is needed to enable efficient evaluation and optimization of scheduling strategies. In the following section, we propose our model of the system.

\subsection{System Model}

We model the system using the generative Markov framework introduced in~\cite{lapkovskis2026brief}, where states $\mathbf{s}_t$ evolve over time according to the Markov property. The framework provides a unified representation of system components and allows their distributions to be modeled, modified, and extended independently. Its sparse factorization makes the high-dimensional model tractable for simulation, inference, and policy learning.
We consider a system consisting of a dedicated server \(v_0\), a set of users \(i \in \mathcal{I}\), and a set of nodes \(v \in V = \{v_0\} \cup \mathcal{I}\), each providing resources \(r \in \mathcal{R}\). The system supports \(|\mathcal{K}|\) types of inference tasks, where each task type \(k \in \mathcal{K}\) consists of a set of subtasks \(p \in \mathcal{P}_k\). Users can both request inference tasks and contribute resources to execute subtasks. The system is controlled by two decision-making policies: a scheduler policy \(\pi\) and an executor policy \(\varsigma\). Together, these policies generate the joint action \(\mathbf{u}_t = \langle \mathbf{u}_t^\pi, \mathbf{u}_t^\varsigma \rangle\).
We assume a fixed interval of 1\,s between timesteps $t$.
We define the system state \(\mathbf{s}_t\) as a tuple of high-dimensional variables:
\begin{equation}
    \mathbf{s}_t = \langle \mathbf{o}_t, \mathbf{q}_t, \mathbf{a}_t, \mathbf{x}_t, \mathbf{y}_t, \mathbf{c}_t, \mathbf{d}_t, \mathbf{u}_t \rangle,
\end{equation}
where each variable captures a specific aspect of the system state. 
Each variable is indexed by the system entities to which it applies, with one dimension assigned to each entity type.
Following the Markov property, variables may depend on other variables within the same or previous time step. These dependencies are generally sparse, since many entities are independent in various state variables. In the following sections, we describe the system variables and the implementations of the decision-making policies.

\subsection{System Variables}

\textbf{Availability states ($\mathbf{o}_t$).}
We represent user availability with the online indicator \(\mathbf{o}_t \in \{0,1\}^{|\mathcal{I}|}\) and the state-duration variable \(\mathbf{d}^o_t \in \mathbb{N}^{|\mathcal{I}|}\), where \(o_{t,i}=1\) indicates that user \(i \in \mathcal{I}\) is online and \(d^o_{t,i}\) denotes the number of time steps since the last online/offline transition. These variables evolve as \(\mathbf{d}^o_t \sim P(\cdot \mid \mathbf{d}^o_{t-1}, \mathbf{o}_{t-1})\) and \(\mathbf{o}_t \sim P(\cdot \mid \mathbf{o}_{t-1}, \mathbf{d}^o_t)\), factorized over users under independent availability processes. The duration dependence captures the fact that transition probabilities change with the time spent in the current state, while \(\mathbf{o}_{t-1}\) selects the corresponding online or offline duration model. For each user, \(d^o_{t,i}\) increments if the state persists and resets to \(0\) if it changes; \(o_{t,i}\) then remains equal to \(o_{t-1,i}\) when \(d^o_{t,i}>0\) and switches to \(1-o_{t-1,i}\) otherwise.

We model availability durations with parametric survival distributions, following findings from~\cite{javadi2011discovering}, and sample transitions from their discrete-time hazards.
We use a Weibull survival function \(S(t) \triangleq \exp(-(t/\lambda)^k)\) for online durations, with \(\lambda=1133.1\) and \(k=0.5764\), and a log-normal survival function \(S(t) \triangleq 1-\mathrm{\Phi}((\log t-\mu)/\sigma)\) for offline durations, with \(\mu=\log(5259)\), \(\sigma=1.5\), and standard normal CDF \(\mathrm{\Phi}(\cdot)\). Given duration \(d-1\), the current state persists with probability \(S(d)/S(d-1)\) and changes otherwise. These parameters yield median online and offline durations of approximately \(10\) minutes and \(1.5\) hours, respectively, and a steady-state user availability of approximately \(10\%\).

\textbf{Inference requests ($\mathbf{q}_t$).}
We represent inference requests with \(\mathbf{q}_t \in \mathcal{Q}^{|\mathcal{I}|}\), where \(\mathcal{Q}=\mathcal{K}\cup\{k_0\}\) and \(k_0\) denotes no active request, and with the request-state duration \(\mathbf{d}^q_t \in \mathbb{N}^{|\mathcal{I}|}\), where \(d^q_{t,i}\) counts the time steps since the last request-state change of user \(i\). The variables evolve as \(\mathbf{d}^q_t \sim P(\cdot \mid \mathbf{d}^q_{t-1},\mathbf{q}_{t-1},\mathbf{o}_t,\mathbf{y}_{t-1})\) and \(\mathbf{q}_t \sim P(\cdot \mid \mathbf{q}_{t-1},\mathbf{o}_t,\mathbf{d}^q_t)\), factorized over users. Users can submit requests only while online; hence, if \(o_{t,i}=0\), we set \(q_{t,i}=k_0\) and \(d^q_{t,i}=0\). Each user has at most one active request, so requests alternate between \(k_0\) and some \(k\in\mathcal{K}\). The duration \(d^q_{t,i}\) increments while the request state persists and resets when a request starts, an active request completes according to \(\mathbf{y}_{t-1}\), or the user abandons it.

In the implementation, we use a Pareto survival function \(S(t) \triangleq 1\) for \(t<\kappa\) and \(S(t) \triangleq (\kappa/t)^\alpha\) for \(t\geq\kappa\), following heavy-tailed inter-request times observed in interactive systems~\cite{benevenuto2009characterizing,crovella1997selfsimilarity}. We use this distribution both for user think time, when \(q_{t-1,i}=k_0\), and patience, when \(q_{t-1,i}\neq k_0\) and the request has not completed. Given duration \(d-1\), the current request state persists with probability \(S(d)/S(d-1)\) and changes otherwise; when a new request starts, we sample its type $k$ from \(\mathrm{Categorical}(p_1,\ldots,p_{|\mathcal{K}|})\). We set \(\kappa=30\,\mathrm{s}\) and \(\alpha=1.5\), yielding a minimum think time and patience of \(30\,\mathrm{s}\), a median of approximately \(48\,\mathrm{s}\), and equal steady-state fractions of time with and without an active request.

\textbf{Allocated resources ($\mathbf{a}_t$).}
We represent allocated resources with \(\mathbf{a}_t \in \mathbb{R}_{\geq 0}^{|\mathcal{R}|\times |V|}\), where \(a_{t,r,v}\) denotes the amount of resource \(r \in \mathcal{R}\) allocated to the system on node \(v \in V\). The variable evolves as \(\mathbf{a}_t \sim P(\cdot \mid \tilde{\mathbf{a}}_{t-1},\mathbf{o}_t)\), where \(\tilde{\mathbf{a}}_{t-1}=[\mathbf{a}_{t-1},\ldots,\mathbf{a}_{t-H}]\) captures the autocorrelation of resource trajectories. We factorize the transition over resources and nodes, condition user resources on availability, and treat the server \(v_0\) as always online. Thus, we set \(a_{t,r,i}=0\) for offline users and sample allocations for online users and the server. Each \(a_{t,r,v}\) represents the total resource capacity allocated to the system, including both used and unused capacity.

In the implementation, we use \(\mathcal{R}{=}\{\texttt{band\_in},\texttt{band\_out},\texttt{cpu},\texttt{memory},\texttt{storage}\}\) and fit one generative ARMA model~\cite{box2015time} per resource. We collect workstation measurements every \(3\,\mathrm{s}\) for \(4\) hours, resample them to \(1\,\mathrm{s}\), standardize each resource, and apply a Yeo--Johnson transformation~\cite{yeo2000transform}. For each \(r\), we fit ARMA models with parameters \(p\in\{1,2,5,10\}\) and \(q\in\{0,1,2,5,10\}\) by maximum likelihood, selecting \(\mathrm{ARMA}^{A,r}_*\) with the lowest AIC~\cite{anderson2004model}. During simulation, we sample from \(\mathrm{ARMA}^{A,r}_*\), invert the transformation, rescale using node-specific mean \(\mu^A_{r,v}\) and standard deviation \(\sigma^A_{r,v}\), and clip to \([\underline{a}^o_{r,v},\overline{a}^o_{r,v}]\), where \(\underline{a}^o_{r,v}=0\) and \(\overline{a}^o_{r,v}\) denote the minimum and maximum permitted allocations. We assume homogeneous user nodes but allow different server parameters.

\textbf{Subtask readiness states ($\mathbf{x}_t$).}
We represent subtask-data readiness with an auxiliary state \(\hat{\mathbf{x}}_t \in \mathcal{X}^{|V|\times|\mathcal{K}|\times P_{\max}}\), a duration variable \(\mathbf{d}^x_t \in \mathbb{N}^{|V|\times|\mathcal{K}|\times P_{\max}}\), and the actual readiness state \(\mathbf{x}_t \in \mathcal{X}^{|V|\times|\mathcal{K}|\times P_{\max}}\), where $\mathcal{X}=\{\texttt{empty},\allowbreak\texttt{download},\allowbreak\texttt{pause},\allowbreak\texttt{ready}\}$, \(P_{\max}=\max_{k\in\mathcal{K}}|\mathcal{P}_k|\), and \(p\in\mathcal{P}_k\) indexes subtasks of task type \(k\). These variables evolve as \(\hat{\mathbf{x}}_t \sim P(\cdot \mid \mathbf{x}_{t-1},\mathbf{o}_t,\mathbf{u}^{x}_{t-1})\), \(\mathbf{d}^x_t \sim P(\cdot \mid \mathbf{d}^x_{t-1},\hat{\mathbf{x}}_t,\mathbf{c}^x_t)\), and \(\mathbf{x}_t \sim P(\cdot \mid \hat{\mathbf{x}}_t,\mathbf{d}^x_t)\), factorized over nodes, task types, and subtasks. The auxiliary state first applies scheduler and executor actions \(\mathbf{u}^{x}_{t-1}=\langle \mathbf{u}^{x,\pi}_{t-1},\mathbf{u}^{x,\varsigma}_{t-1}\rangle\): the scheduler can start or cancel downloads, while the executor can pause, resume, or abort them. We suppress transitions for offline users and treat the server \(v_0\) as always online; moreover, the server stores all subtask data a priori.

The duration \(d^x_{t,v,k,p}\) matters only when \(\hat{x}_{t,v,k,p}=\texttt{download}\). For all other states, we set it to \(0\). During an active download, \(d^x_{t,v,k,p}\) increments while the accumulated storage consumption \(c^x_{t,\texttt{storage},v,k,p}\) remains below the required subtask-data size \(c_{k,p}\), and resets to \(0\) once \(c^x_{t,\texttt{storage},v,k,p}=c_{k,p}\). The actual state \(\mathbf{x}_t\) copies \(\hat{\mathbf{x}}_t\), except that \(\hat{x}_{t,v,k,p}=\texttt{download}\) becomes \(x_{t,v,k,p}=\texttt{ready}\) when \(d^x_{t,v,k,p}=0\), indicating download completion. In the implementation, paused downloads retain their accumulated storage, ready subtasks keep storage fixed at \(c_{k,p}\), and all non-storage download consumptions vanish outside the \texttt{download} state.

\textbf{Subtask execution states ($\mathbf{y}_t$).}
We represent subtask execution with an auxiliary state \(\hat{\mathbf{y}}_t \in \mathcal{Y}^{|\mathcal{I}|\times |V|\times |\mathcal{K}|\times P_{\max}}\), a duration variable \(\mathbf{d}^y_t \in \mathbb{N}^{|\mathcal{I}|\times |V|\times |\mathcal{K}|\times P_{\max}}\), and the actual execution state \(\mathbf{y}_t \in \mathcal{Y}^{|\mathcal{I}|\times |V|\times |\mathcal{K}|\times P_{\max}}\), where \(\mathcal{Y}=\{\texttt{idle}, \allowbreak \texttt{download}, \allowbreak \texttt{execution}, \allowbreak \texttt{upload}, \allowbreak \texttt{done}\}\). The entry \(y_{t,i,v,k,p}\) describes the execution state of subtask \(p\in\mathcal{P}_k\) of user \(i\)'s request of type \(k\) on node \(v\). These variables evolve as \(\hat{\mathbf{y}}_t \sim P(\cdot \mid \mathbf{y}_{t-1},\mathbf{o}_t,\mathbf{u}^{y}_{t-1})\), \(\mathbf{d}^y_t \sim P(\cdot \mid \mathbf{d}^y_{t-1},\hat{\mathbf{y}}_t,\mathbf{c}^y_t)\), and \(\mathbf{y}_t \sim P(\cdot \mid \hat{\mathbf{y}}_t,\mathbf{d}^y_t)\), factorized over users, nodes, task types, and subtasks. The scheduler action \(\mathbf{u}^{y,\pi}_{t-1}\) assigns, cancels, or finalizes subtasks, while the executor action \(\mathbf{u}^{y,\varsigma}_{t-1}\) can abort running subtasks. Assigning a subtask starts with \texttt{download} on user nodes and directly with \texttt{execution} on the server \(v_0\); offline user nodes reset non-completed subtasks to \texttt{idle}, whereas \texttt{done} subtasks can still be finalized because the server already has their outputs.

The duration \(d^y_{t,i,v,k,p}\) tracks the current active phase \(y\in\tilde{\mathcal{Y}}=\{\texttt{download}, \allowbreak \texttt{execution}, \allowbreak \texttt{upload}\}\). It increments while the phase continues and resets to \(0\) when the phase completes, is interrupted, or reaches \texttt{done}. The actual state \(\mathbf{y}_t\) copies \(\hat{\mathbf{y}}_t\), except that duration resets advance active phases as \(\texttt{download}\rightarrow\texttt{execution}\), \(\texttt{execution}\rightarrow\texttt{upload}\) and \(\texttt{upload}\rightarrow\texttt{done}\) on user nodes, and \(\texttt{execution}\rightarrow\texttt{done}\) on the server. In the implementation, we model phase completion with discrete-time hazard models~\cite{singer2003applied} fitted separately for each \(y\in\tilde{\mathcal{Y}}\). We fit these models from measured executions of representative computer-vision subtasks under different bandwidth levels, batch sizes \(\{1,2,5,10,20,50\}\), and subtask types. The covariates include standardized resource consumptions \(\mathbf{c}^y_t\), batch size, \(\log d^y_{t,i,v,k,p}\), polynomial interactions up to degree \(q\in\{1,2,3\}\), and one-hot subtask-type indicators; we select a common degree \(q\) by average AIC across phases. During simulation, subtasks with the same node, subtask type, and execution age form a batch, share one sampled completion event, and either advance by resetting \(d^y\) to \(0\) or continue with \(d^y+1\).

\textbf{Resource consumption ($\mathbf{c}_t$).}
We define total resource consumption as \(\mathbf{c}_t=\langle\mathbf{c}^x_t,\mathbf{c}^y_t\rangle\), where \(\mathbf{c}^x_t \in \mathbb{R}_{\geq 0}^{|\mathcal{R}^x|\times |\mathcal{V}|\times |\mathcal{K}|\times P_{\max}}\) captures subtask-data preparation and \(\mathbf{c}^y_t \in \mathbb{R}_{\geq 0}^{|\mathcal{R}^y|\times |\mathcal{I}|\times |\mathcal{V}|\times |\mathcal{K}|\times P_{\max}}\) captures request execution. We use \(\mathcal{R}^x=\mathcal{R}\setminus\{\texttt{band\_out}\}\) and \(\mathcal{R}^y=\mathcal{R}\setminus\{\texttt{storage}\}\), with execution-stage subsets for downloading, computation, and uploading. The variables evolve jointly as \((\mathbf{c}^x_t,\mathbf{c}^y_t)\sim P(\cdot\mid \tilde{\mathbf{c}}^x_{t-1},\tilde{\mathbf{c}}^y_{t-1},\mathbf{a}_t,\hat{\mathbf{x}}_t,\hat{\mathbf{y}}_t,\mathbf{d}^x_{t-1},\mathbf{d}^y_{t-1})\), where the history windows \(\tilde{\mathbf{c}}^x_{t-1}\) and \(\tilde{\mathbf{c}}^y_{t-1}\) capture autocorrelation. We condition on allocated resources because they bound feasible consumption, on auxiliary states because inactive subtasks consume no resources, and on durations because subtasks with the same phase age may share resources through batching. Server and user consumptions are coupled: user transfers consume server bandwidth, while limited server bandwidth constrains user-side transfer rates.

In the implementation, we fit generative ARMA models to measured consumption traces after resampling to \(1\,\mathrm{s}\), standardization, and Yeo--Johnson transformation. For \(\mathbf{c}^x_t\), we collect traces by downloading subtask data in \(1\,\mathrm{Kb}\) chunks under representative inbound bandwidth levels obtained by applying Lloyd--Max quantization~\cite{lloyd1982quant} to the inbound-bandwidth traces collected for allocated-resource modeling. We model storage as a delta and accumulate it over time. For \(\mathbf{c}^y_t\), we collect traces from representative 
subtasks under different bandwidth levels, batch sizes \(\{1,2,5,10,20,50\}\), and subtask types. For each resource and stage, we fit ARMA candidates with \(p\in\{1,2,5,10\}\) and \(q\in\{0,1,2,5,10\}\), selecting the model with the lowest AIC. During simulation, we sample node-level consumption, interpolate scaling statistics over the relevant bandwidth or batch-size levels using inverse-distance weighting~\cite{shawky2025comparative}, clip transfer consumption to the effective bandwidth shared with the server, proportionally downscale CPU-dependent throughput when CPU is overcommitted, and divide node-level consumption among concurrent transfers or same-age execution batches.

\subsection{Decision-Making Policies}
The system uses two types of policies. The server-side \emph{scheduler} \(\pi\) issues readiness and execution actions, \(\mathbf{u}^{x,\pi}_t\) and \(\mathbf{u}^{y,\pi}_t\), while the node-side \emph{executor} \(\varsigma\) issues constraint-enforcement actions, \(\mathbf{u}^{x,\varsigma}_t\) and \(\mathbf{u}^{y,\varsigma}_t\). Their action domains are \(\mathcal{U}_{x,\pi}=\{\texttt{none}, \allowbreak \texttt{download}, \allowbreak \texttt{cancel}\}\), \(\mathcal{U}_{y,\pi}=\{\texttt{none}, \allowbreak \texttt{execute}, \allowbreak \texttt{finish}, \allowbreak \texttt{cancel}\}\), \(\mathcal{U}_{x,\varsigma}{=}\{\texttt{none}, \allowbreak \texttt{abort}, \allowbreak \texttt{pause}, \allowbreak \texttt{resume}\}\), and \(\mathcal{U}_{y,\varsigma}=\{\texttt{none}, \allowbreak \texttt{abort}\}\). Both policies condition on the current state excluding the actions being generated, i.e., \(\mathbf{u}^{\pi}_t\sim\pi(\cdot\mid\mathbf{s}_t\setminus\mathbf{u}_t)\) and \(\mathbf{u}^{\varsigma}_t\sim\varsigma(\cdot\mid\mathbf{s}_t\setminus\mathbf{u}_t)\). Executor actions take precedence over scheduler actions.

\textbf{Executor policy.}
The executor policy \(\varsigma\) enforces local feasibility by comparing total consumption from \(\mathbf{c}^x_t\) and \(\mathbf{c}^y_t\) with allocated resources \(\mathbf{a}_t\). Under non-storage overflow, it pauses all active subtask-data downloads on the affected node, since they share a single download channel and can later resume. Under storage overflow, it aborts downloads with the largest storage footprints until feasibility is restored. It also aborts executions whose required subtask data are no longer available or have been aborted. If overflow remains, it aborts executions in the order \(\texttt{download}\rightarrow\texttt{execution}\rightarrow\texttt{upload}\); for \texttt{execution}, it groups subtasks by node, subtask, and execution age, and greedily aborts the heaviest normalized-load batches. Paused downloads resume once the node has no active executions and no non-storage overflow.

\textbf{Centralized scheduler policy.}
The centralized scheduler represents fully centralized inference baseline. Since all executions run on the server \(v_0\), which stores all subtask data a priori, it schedules no user-node downloads. For each active request \(q_{t,i}=k\), it cancels obsolete executions, finishes requests whose required subtasks have reached \texttt{done}, and assigns missing subtasks \(p\in\mathcal{P}_k\) only to \(v_0\), subject to available server memory estimated by \(\widehat{m}^y_{k,p,n}\).

\textbf{Uniform scheduler policy.}
The uniform scheduler uses online volunteered nodes opportunistically. For each online node, the policy schedules downloading at most one missing subtask data whose storage requirement and conservative download-memory estimate \(\widehat{m}^x\) fit within the node's free capacity; the server is excluded because it already stores all subtask data. The policy cancels obsolete executions, finishes completed requests, identifies requested but unplaced user--subtask pairs, and assigns them uniformly at random among feasible nodes. A node is feasible if it is online, has \(x_{t,v,k,p}=\texttt{ready}\), and has enough free memory for the marginal execution-memory estimate \(\widehat{m}^y_{k,p,n}\). We precompute \(\widehat{m}^x\) and \(\widehat{m}^y_{k,p,n}\) from learned memory models using a \(\mu+2\sigma\) safety margin, interpolating \(\widehat{m}^y_{k,p,n}\) across measured batch sizes. Memory estimates are needed because memory, unlike CPU and bandwidth, is a hard constraint.

\textbf{Affinity-based scheduler policy.}
The affinity-based scheduler uses the same download, cancellation, and finishing logic as the uniform policy, but assigns executions to improve batching. For each required subtask \((k,p)\), it prioritizes eligible nodes by the number of active executions of the same subtask already placed on them, excluding offline nodes and nodes without ready subtask data. Using \(\widehat{m}^y_{k,p,n}\), it computes how many additional executions fit in each eligible node's free memory and fills the resulting assignment slots with randomly shuffled users requiring \((k,p)\). This greedily co-locates same-subtask executions to increase batch sizes while aiming to preserve memory feasibility.

\subsection{Optimization Problems} \label{sec:problems}

Given the proposed system model, two optimization problems naturally arise. First, for a fixed executor policy \(\varsigma\), one can optimize the scheduler policy \(\pi\) to maximize QoS, e.g., by minimizing long-term average request latency:
\begin{equation}
\max_\pi\; \liminf_{T\rightarrow\infty}\,\mathbb{E}\left[\frac{1}{T} \sum_{t=1}^T r(\mathbf{s}_t) \mid \mathbf{s}_0, \mathbf{u}^\pi_{0:t-1} \sim \pi \right],
\end{equation}
where $r(\mathbf{s}_t) \triangleq -\mathbf{1}^\top\boldsymbol{\delta}(\mathbf{q}_t, \mathbf{d}^q_t) $ is a reward function, and $\boldsymbol{\delta}(\mathbf{q}_t, \mathbf{d}^q_t) \in \{0,1\}^{|\mathcal{I}|}$ denotes per-user latency increments.

Second, for a fixed scheduler \(\pi\) and executor policy \(\varsigma\), one can optimize the allocation of dedicated server resources. In this case, the objective is to determine the amount of dedicated capacity required to satisfy a target QoS constraint under the given model and policies; e.g., one may require request latency to remain below \(10\,\mathrm{s}\) for \(99\%\) of requests, corresponding to P99 latency. We define this problem as
\begin{equation}
    \begin{aligned}
        \min_{\mathbf{a}_{\dagger}}&\; \lVert \mathbf{a}_{\dagger} \rVert_{\mathbf{w}} \\
        \mathrm{s.t.}& \quad \mathbb{P}\!\left(d^q_{t,i} \le \tau_k \mid q_{t+1,i} = k_0,\, q_{t,i} = k\right) \ge 1-\epsilon_k, \quad \forall k\in\mathcal{K},
    \end{aligned}
\end{equation}
where $\mathbf{a}_{\dagger}$ is a nominal, i.e., maximum, server resource capacity that parametrizes transition probabilities of $\mathbf{a}_t$, and $\lVert \cdot \rVert_{\mathbf{w}}$ is a weighted $\ell^1$ norm with per-resource cost weights $\mathbf{w}$, $\tau_k$ are per-request-type latency thresholds, and $\epsilon_k \in (0,1)$ are significance levels. Formal definitions of these problems are provided in~\cite{lapkovskis2026brief}.

It is noteworthy to mention that this paper does not focus on designing a scheduling algorithm or resource-allocation optimizer. Instead, our primary focus is to define the collaborative distributed inference system and develop a model of its dynamics. We use this model to simulate fixed policies under different configurations and compare collaborative distributed inference with centralized inference. Future work will use this model to address the optimization problems.

\section{Evaluation}

\subsection{Experimental Setup}

We evaluate the proposed system by simulating trajectories of the generative model, sampling \(\mathbf{s}_t \sim P(\cdot \mid \mathbf{s}_{t-1})\). We parameterize ARMA resource-consumption and hazard-based duration models from measurements collected on a MacBook M4 Pro with \(24\,\mathrm{GB}\) RAM and \(512\,\mathrm{GB}\) storage. We consider four task types: \emph{light} and \emph{heavy}, each with \(5\) or \(10\) subtasks, sampled with probabilities \(\{0.4, \allowbreak 0.3, \allowbreak 0.2, \allowbreak 0.1\}\) to reflect that simpler tasks are typically requested more often. Subtasks are instantiated using established computer-vision models. Each simulation runs for \(100{,}000\) time steps, or approximately \(28\) hours. We vary user count \(\{100,1{,}000,5{,}000,10{,}000\}\), user capacity \(\{1\times,2\times\}\) of \emph{base resource profile}, server capacity \(\{10\times, \allowbreak 30\times, \allowbreak 50\times, \allowbreak 80\times, \allowbreak 100\times\}\), and scheduler policy. These dimensions test how the system behaves as demand and volunteered resources grow together, how offloading depends on user-device capacity, how much dedicated capacity is needed to sustain QoS, and how scheduling affects system behavior.

We define the \emph{base resource profile} as a moderate, non-intrusive contribution from a user device:
\(100\,\mathrm{Mbit}\) inbound/outbound bandwidth, \(3\,\mathrm{GB}\) storage, \(2.5\,\mathrm{GB}\) memory, and \(200\%\) CPU. The bandwidth represents neither a weak nor exceptional connection; the storage can hold several small subtasks or one large subtask with a few smaller ones. We set memory and CPU to rounded values slightly above the largest measured average subtask consumption at batch size \(2\), making small batches of demanding subtasks feasible with high probability and larger lightweight batches more likely. These values define the \emph{maximum allocation thresholds} \(\overline{a}^{o}_{r,v}\), while realized capacities \(\mathbf{a}_t\) still vary over time. We report request completions and cancellations, P99 latency as the primary QoS metric, dedicated resource consumption, and server/user subtask completion rates. We focus on P99 latency because other latency summaries, such as mean and P95 latency, showed similar trends and are omitted due to space constraints.

For full reproducibility, we published our implementation code on GitHub.\footnote{https://anonymous.4open.science/r/AutoscalingArchitecture-C41F}

\subsection{Results}

\begin{figure} [!t]
    \centering
    \begin{subfigure}[t]{0.19\textwidth}
        \centering
        \includegraphics[width=\textwidth]{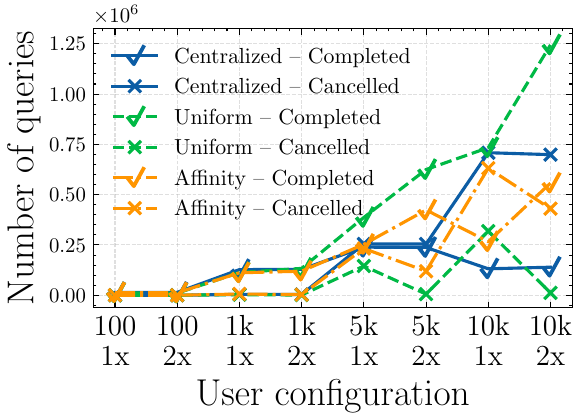}
        \caption{Server $10\times$}
        \label{fig:completed_queries_10x}
    \end{subfigure}
    \hfill
    \begin{subfigure}[t]{0.19\textwidth}
        \centering
        \includegraphics[width=\textwidth]{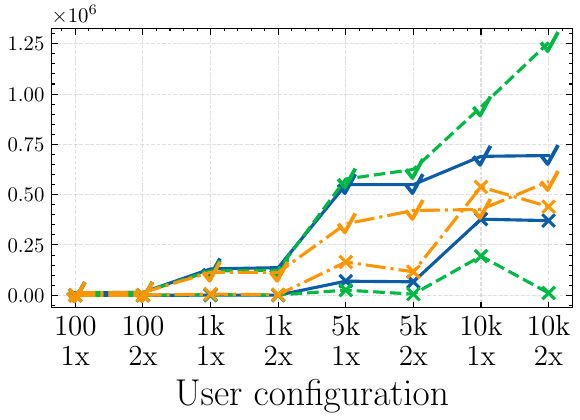}
        \caption{Server $30\times$}
        \label{fig:completed_queries_30x}
    \end{subfigure}
    \hfill
    \begin{subfigure}[t]{0.19\textwidth}
        \centering
        \includegraphics[width=\textwidth]{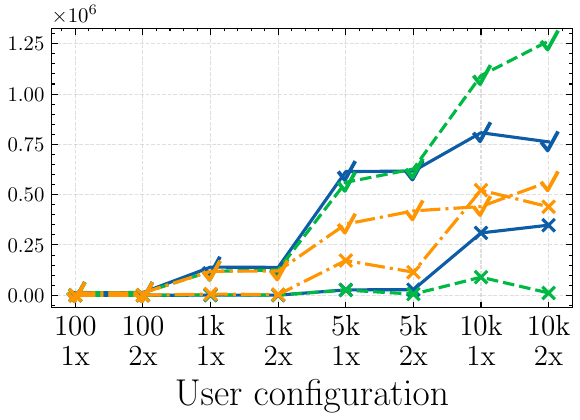}
        \caption{Server $50\times$}
        \label{fig:completed_queries_50x}
    \end{subfigure}
    \hfill
    \begin{subfigure}[t]{0.19\textwidth}
        \centering
        \includegraphics[width=\textwidth]{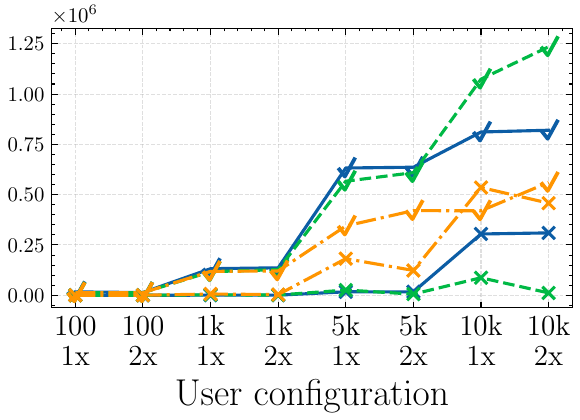}
        \caption{Server $80\times$}
        \label{fig:completed_queries_80x}
    \end{subfigure}
    \hfill
    \begin{subfigure}[t]{0.19\textwidth}
        \centering
        \includegraphics[width=\textwidth]{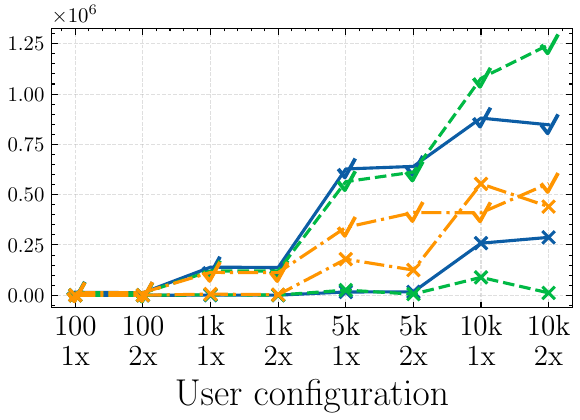}
        \caption{Server $100\times$}
        \label{fig:completed_queries_100x}
    \end{subfigure}

    \caption{Completed vs. canceled requests under different scheduler policies and server capacities. Each subfigure corresponds to a different server capacity multiplier. The \(x\)-axis varies the user configuration, i.e., the number of users ($|\mathcal{I}|$) and the per-user resource-capacity multiplier.}

    \label{fig:completed_queries}
\end{figure}

\textbf{Completed vs. canceled tasks.}
Fig.~\ref{fig:completed_queries} compares completed and canceled requests. The centralized policy is unaffected by user capacity because all subtasks run on the server. Under the centralized policy, at \(10\times\) capacity, completions grow only through \(1{,}000\) users, then plateau and decline as cancellations surge. Higher capacities delay saturation, but gains diminish beyond \(30\times\); at large user counts, cancellations still equal roughly one third to one half of completions. 

The uniform policy matches the centralized through \(1{,}000\) users and improves with \(2\times\) user capacity. At \(5{,}000\) users, it performs better at \(10\times\) and \(30\times\) server capacity and comparably above them. At \(10{,}000\), it yields more completions and fewer cancellations at every server capacity, including \(100\times\). Thus, the growing volunteered resource pool offsets demand more effectively than centralized scaling, whose gains from \(50\times\) to \(100\times\) are incremental. Centralized execution may also form larger batches that save resources but delay individual subtasks, whereas uniform distribution creates smaller batches across more nodes. 

The affinity-based policy is competitive through \(1{,}000\) users and beats the centralized policy at \(10\times\) capacity, but otherwise generally yields the worst completion-to-cancellation balance. Its larger batches may extend execution and increase cancellation risk; node departures or executor aborts can also interrupt several batched subtasks simultaneously. Overall, distributed inference is effective and feasible, but batching efficiency must be balanced against latency and failure risk, motivating scheduler optimization using the proposed model.

\begin{figure} [!t]
    \centering
    \begin{subfigure}[t]{0.19\textwidth}
        \centering
        \includegraphics[width=\textwidth]{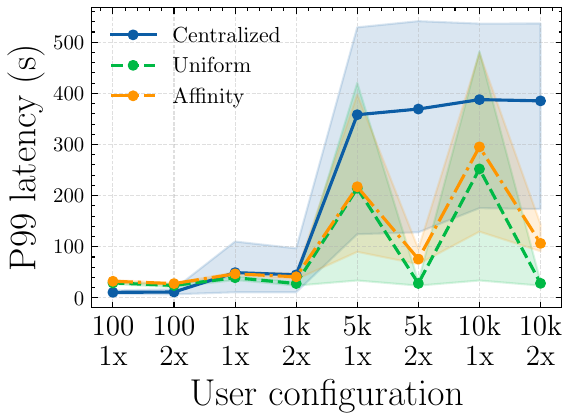}
        \caption{Server $10\times$}
        \label{fig:p99_latencies_10x}
    \end{subfigure}
    \hfill
    \begin{subfigure}[t]{0.19\textwidth}
        \centering
        \includegraphics[width=\textwidth]{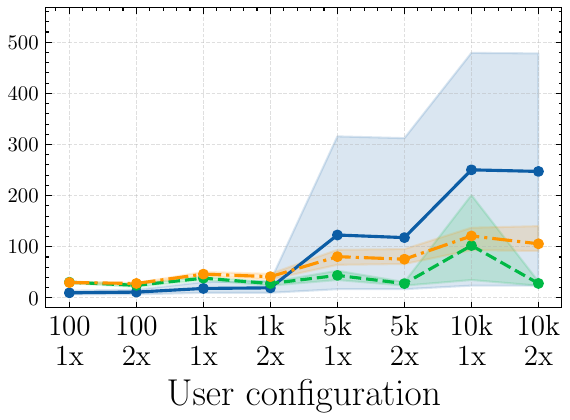}
        \caption{Server $30\times$}
        \label{fig:p99_latencies_30x}
    \end{subfigure}
    \hfill
    \begin{subfigure}[t]{0.19\textwidth}
        \centering
        \includegraphics[width=\textwidth]{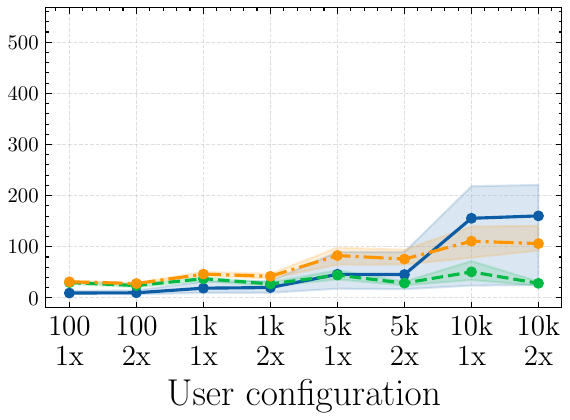}
        \caption{Server $50\times$}
        \label{fig:p99_latencies_50x}
    \end{subfigure}
    \hfill
    \begin{subfigure}[t]{0.19\textwidth}
        \centering
        \includegraphics[width=\textwidth]{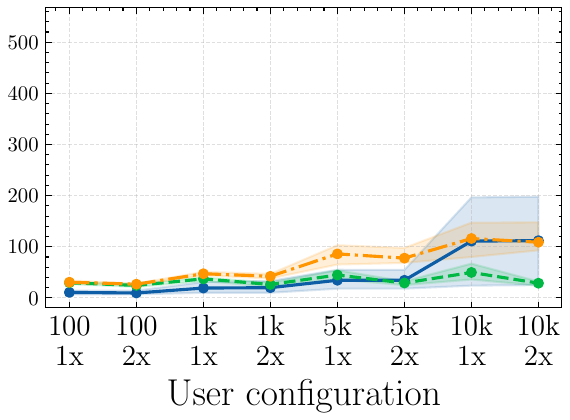}
        \caption{Server $80\times$}
        \label{fig:p99_latencies_80x}
    \end{subfigure}
    \hfill
    \begin{subfigure}[t]{0.19\textwidth}
        \centering
        \includegraphics[width=\textwidth]{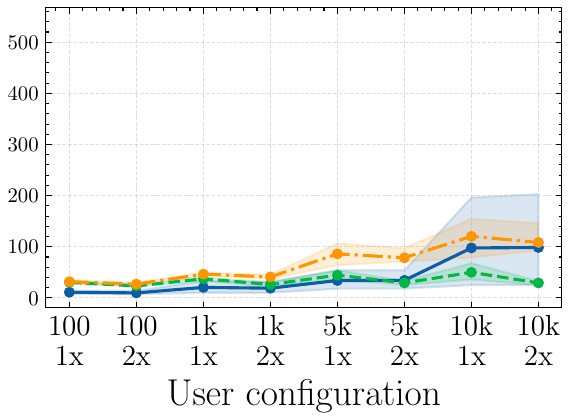}
        \caption{Server $100\times$}
        \label{fig:p99_latencies_100x}
    \end{subfigure}

    \caption{P99 request latency under different scheduler policies and server capacities. Each subfigure corresponds to a different server capacity. The \(x\)-axis shows user configurations. The curves represent average P99 latencies across request types. The faded regions represent the range of P99 latencies across request types, from the lowest to the highest.}

    \label{fig:p99_latencies}
\end{figure}

\textbf{Task latencies.}
Fig.~\ref{fig:p99_latencies} reports P99 request latency. The centralized policy performs best or comparably through \(1{,}000\) users, and at \(5{,}000\) as server capacity grows. At \(10{,}000\), the uniform policy remains substantially better. Centralized latency varies more across request types under higher demand. Besides differences in task complexity, unequal request probabilities lead to frequent subtasks forming larger batches that run longer and reduce the resources available to other requests. More server capacity mitigates but does not eliminate this effect. 

The uniform policy matches the centralized policy through \(1{,}000\) users, outperforms it at \(5{,}000\) with up to \(30\times\) server capacity, and remains comparable above that. At \(10{,}000\), it consistently achieves much lower P99 latency. Its variation under constrained configurations likely arises because weaker users store fewer models, particularly for heavy tasks, forcing more execution onto the server. Increasing user and server capacities distributes more subtasks and largely removes this variation. 

At \(10\times\) server capacity, the affinity-based policy is slightly worse than the uniform policy but better than the centralized policy at large user counts. Beyond \(30\times\), its latency changes little and generally remains above the uniform policy, likely because larger batches delay and couple multiple requests. Overall, both distributed policies show little improvement beyond \(30\times\)--\(50\times\) server capacity, i.e., this range provides sufficient dedicated capacity to complement volunteered resources. Centralized scaling improves slowly and does not surpass the uniform policy at a large scale. Thus, distributed scheduling provides better tail latency and under lower server capacity than centralized scheduling.

\begin{figure}[!t]
    \centering
    \begin{subfigure}[t]{0.19\textwidth}
        \centering
        \includegraphics[width=\textwidth]{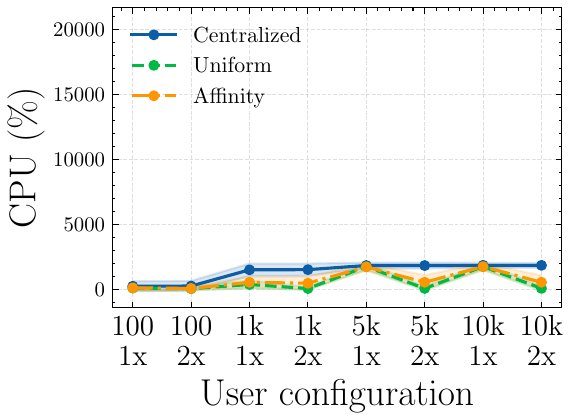}
        \caption{Server $10\times$}
        \label{fig:server_cpu_10x}
    \end{subfigure}
    \hfill
    \begin{subfigure}[t]{0.19\textwidth}
        \centering
        \includegraphics[width=\textwidth]{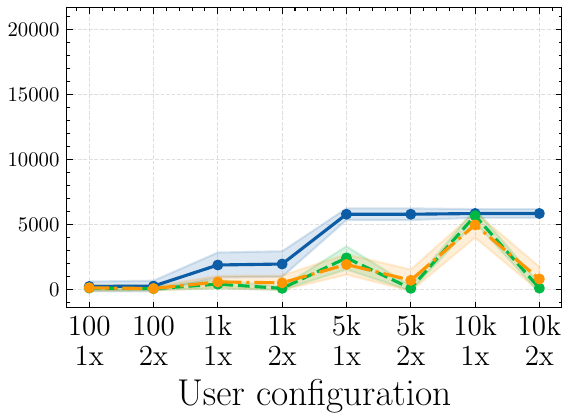}
        \caption{Server $30\times$}
        \label{fig:server_cpu_30x}
    \end{subfigure}
    \hfill
    \begin{subfigure}[t]{0.19\textwidth}
        \centering
        \includegraphics[width=\textwidth]{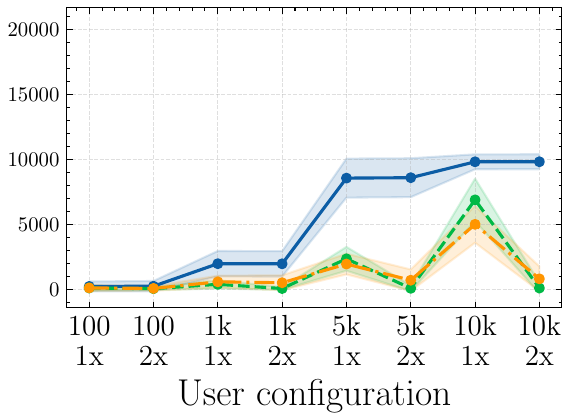}
        \caption{Server $50\times$}
        \label{fig:server_cpu_50x}
    \end{subfigure}
    \hfill
    \begin{subfigure}[t]{0.19\textwidth}
        \centering
        \includegraphics[width=\textwidth]{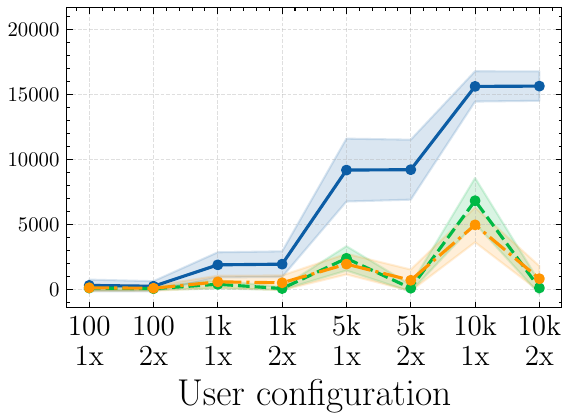}
        \caption{Server $80\times$}
        \label{fig:server_cpu_80x}
    \end{subfigure}
    \hfill
    \begin{subfigure}[t]{0.19\textwidth}
        \centering
        \includegraphics[width=\textwidth]{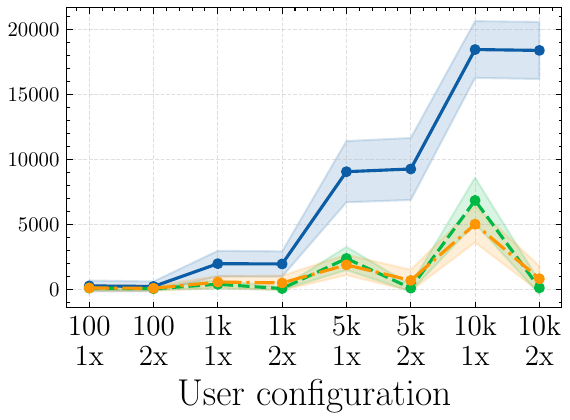}
        \caption{Server $100\times$}
        \label{fig:server_cpu_100x}
    \end{subfigure}
    \hfill
    \begin{subfigure}[t]{0.19\textwidth}
        \centering
        \includegraphics[width=\textwidth]{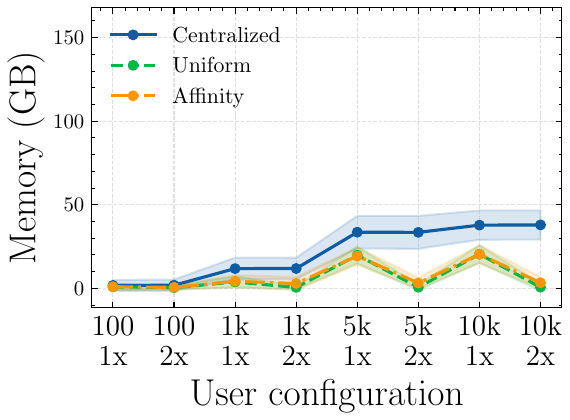}
        \caption{Server $10\times$}
        \label{fig:server_memory_10x}
    \end{subfigure}
    \hfill
    \begin{subfigure}[t]{0.19\textwidth}
        \centering
        \includegraphics[width=\textwidth]{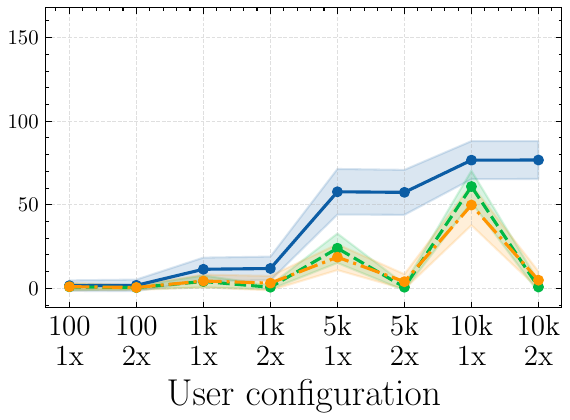}
        \caption{Server $30\times$}
        \label{fig:server_memory_30x}
    \end{subfigure}
    \hfill
    \begin{subfigure}[t]{0.19\textwidth}
        \centering
        \includegraphics[width=\textwidth]{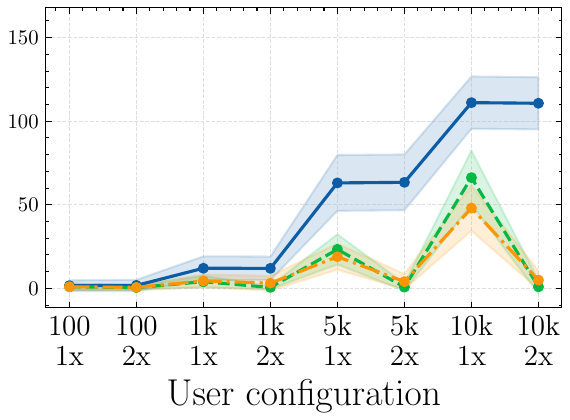}
        \caption{Server $50\times$}
        \label{fig:server_memory_50x}
    \end{subfigure}
    \hfill
    \begin{subfigure}[t]{0.19\textwidth}
        \centering
        \includegraphics[width=\textwidth]{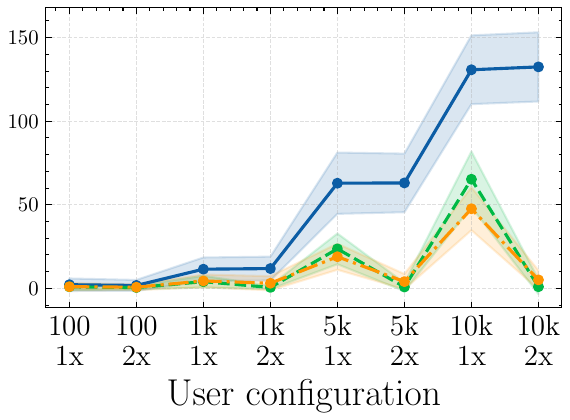}
        \caption{Server $80\times$}
        \label{fig:server_memory_80x}
    \end{subfigure}
    \hfill
    \begin{subfigure}[t]{0.19\textwidth}
        \centering
        \includegraphics[width=\textwidth]{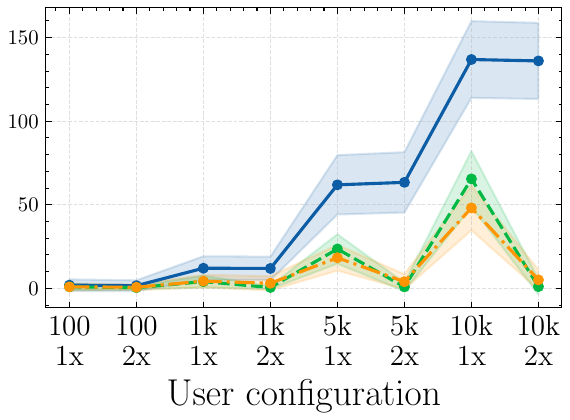}
        \caption{Server $100\times$}
        \label{fig:server_memory_100x}
    \end{subfigure}
    \caption{Server resource consumption under different scheduler policies and server capacities. The first row reports CPU utilization, where \(100\%\) corresponds to full utilization of one CPU core, and the second row reports memory usage. The curves represent average resource consumptions over time. The faded regions represent the corresponding standard deviations.}

    \label{fig:resource_consumption}
\end{figure}

\textbf{Dedicated resource consumption.}
Fig.~\ref{fig:resource_consumption} shows that the centralized policy consistently uses more dedicated resources. CPU consumption reaches every allocation limit, indicating persistent saturation. Since consumption is recorded before the executor resolves hard-constraint violations, mean memory demand reaches nearly twice the allocated \(25\,\mathrm{GB}\) at \(10\times\) capacity and slightly exceeds \(75\,\mathrm{GB}\) at \(30\times\), causing the subtask aborts. At higher capacities, memory remains within its allocation and grows slowly, suggesting that demand approaches its required level while additional CPU accelerates processing, causing slight further increases in memory usage.

Both distributed policies use substantially fewer dedicated resources and never exceed their allocations. Consumption changes little between \(30\times\) and \(100\times\), indicating that approximately \(30\times\)--\(50\times\) capacity sufficiently complements volunteered resources. The affinity-based policy is slightly more efficient than the uniform policy, likely because it considers node loads and exploits batching. Overall, distributed scheduling accommodates growing demand without proportional expansion of dedicated infrastructure.

\begin{figure} [!t]
    \centering
    \begin{subfigure}[t]{0.19\textwidth}
        \centering
        \includegraphics[width=\textwidth]{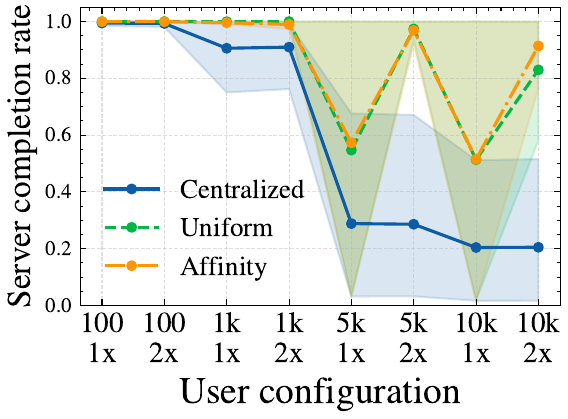}
        \caption{Server $10\times$}
        \label{fig:server_subtask_rate_10x}
    \end{subfigure}
    \hfill
    \begin{subfigure}[t]{0.19\textwidth}
        \centering
        \includegraphics[width=\textwidth]{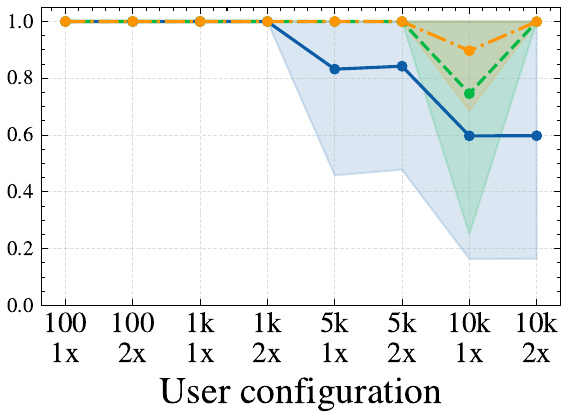}
        \caption{Server $30\times$}
        \label{fig:server_subtask_rate_30x}
    \end{subfigure}
    \hfill
    \begin{subfigure}[t]{0.19\textwidth}
        \centering
        \includegraphics[width=\textwidth]{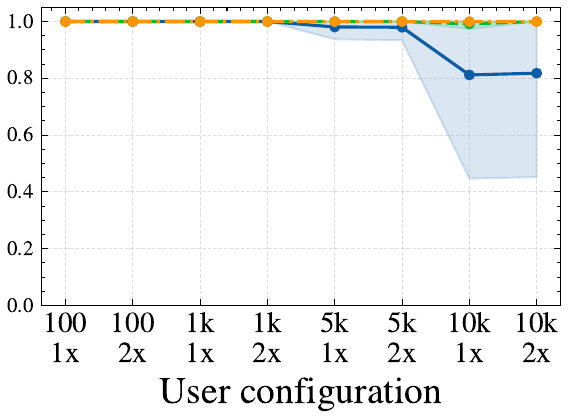}
        \caption{Server $50\times$}
        \label{fig:server_subtask_rate_50x}
    \end{subfigure}
    \hfill
    \begin{subfigure}[t]{0.19\textwidth}
        \centering
        \includegraphics[width=\textwidth]{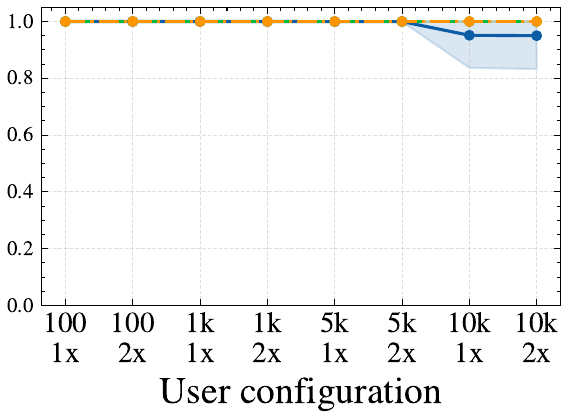}
        \caption{Server $80\times$}
        \label{fig:server_subtask_rate_80x}
    \end{subfigure}
    \hfill
    \begin{subfigure}[t]{0.19\textwidth}
        \centering
        \includegraphics[width=\textwidth]{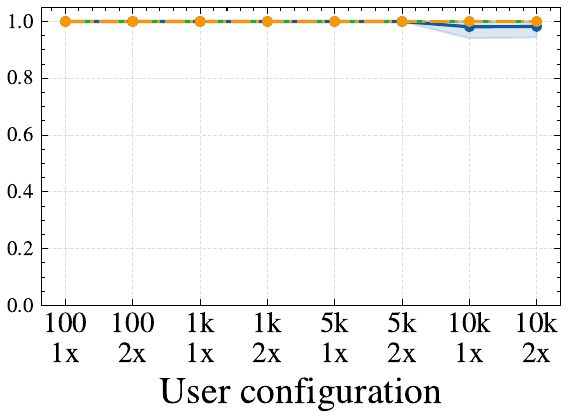}
        \caption{Server $100\times$}
        \label{fig:server_subtask_rate_100x}
    \end{subfigure}
    \hfill
    \begin{subfigure}[t]{0.19\textwidth}
        \centering
        \includegraphics[width=\textwidth]{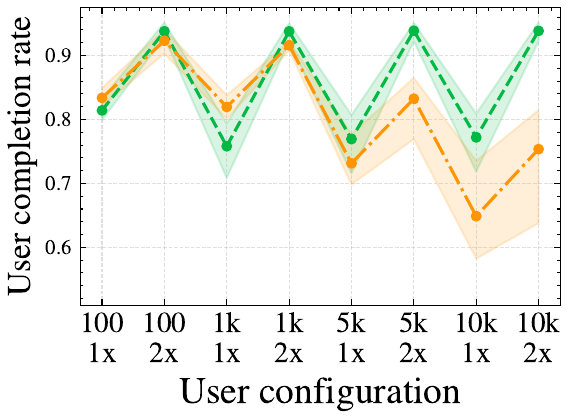}
        \caption{Server $10\times$}
        \label{fig:user_subtask_rate_10x}
    \end{subfigure}
    \hfill
    \begin{subfigure}[t]{0.19\textwidth}
        \centering
        \includegraphics[width=\textwidth]{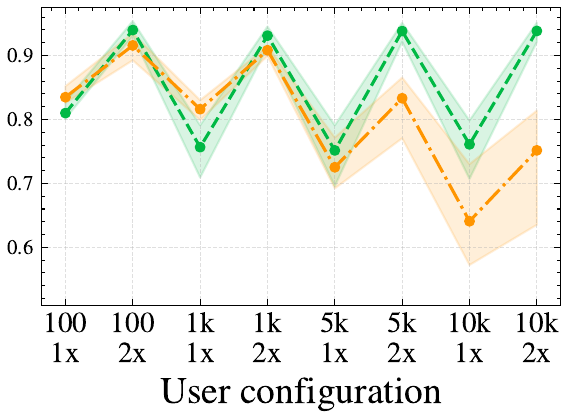}
        \caption{Server $30\times$}
        \label{fig:user_subtask_rate_30x}
    \end{subfigure}
    \hfill
    \begin{subfigure}[t]{0.19\textwidth}
        \centering
        \includegraphics[width=\textwidth]{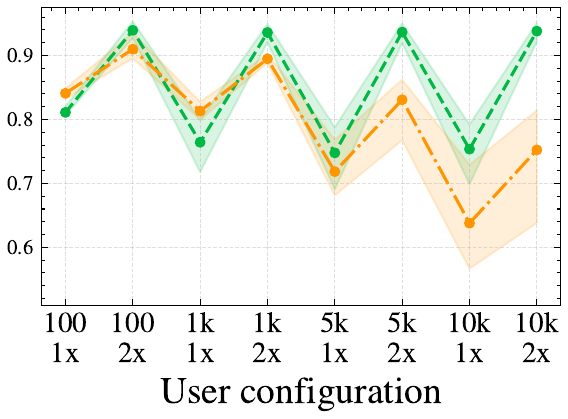}
        \caption{Server $50\times$}
        \label{fig:user_subtask_rate_50x}
    \end{subfigure}
    \hfill
    \begin{subfigure}[t]{0.19\textwidth}
        \centering
        \includegraphics[width=\textwidth]{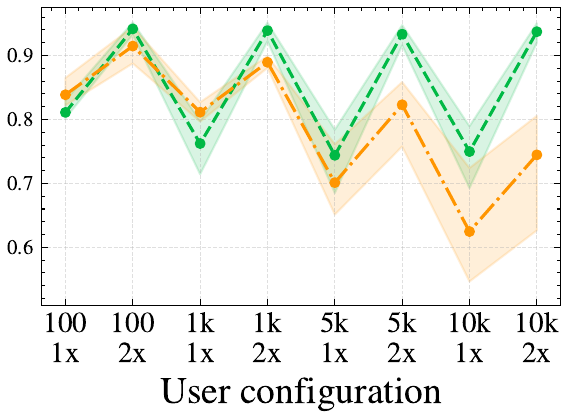}
        \caption{Server $80\times$}
        \label{fig:user_subtask_rate_80x}
    \end{subfigure}
    \hfill
    \begin{subfigure}[t]{0.19\textwidth}
        \centering
        \includegraphics[width=\textwidth]{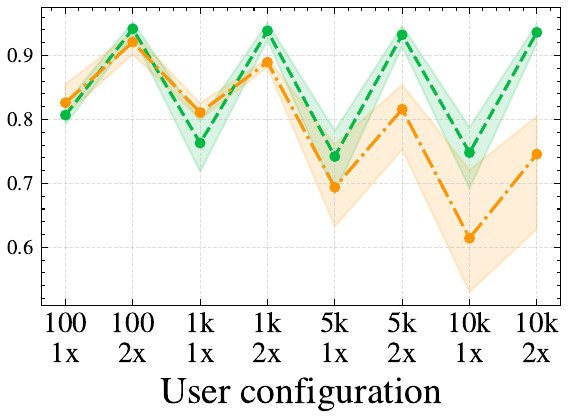}
        \caption{Server $100\times$}
        \label{fig:user_subtask_rate_100x}
    \end{subfigure}

    \caption{Subtask completion rates under different scheduler policies and server capacities. Completion rate denotes the fraction of scheduled subtasks completed successfully. The first and second rows show completion rates on the server and user nodes, respectively.}

    \label{fig:subtask_rates}
\end{figure}

\textbf{Subtask completion.}
Fig.~\ref{fig:subtask_rates} reports subtask completion rates on server and user nodes; the centralized policy is omitted from user-node plots. Under distributed policies, user-side rates are largely independent of server capacity. The uniform policy achieves approximately \(0.75\)--\(0.95\), improving with user capacity because stronger nodes better tolerate resource fluctuations. Affinity-based scheduling performs better at \(100\) and \(1{,}000\) users with \(1\times\) capacity, but degrades as user count grows, especially on weaker nodes. Larger batches may exceed expected resource demands, leading to the simultaneous abort of multiple subtasks due to node departures or resource-constraint violations. 

Centralized server-side rates at \(10\times\) capacity fall from \(0.9\)--\(1.0\) for up to \(1{,}000\) users to approximately \(0.2\) at \(10{,}000\). At \(30\times\), they remain near \(1.0\) through \(1{,}000\) users but decline to approximately \(0.6\) at \(10{,}000\); only \(100\times\) capacity approaches consistently complete execution. Distributed policies reach near-perfect server-side rates with considerably less capacity, particularly for \(2\times\) users, because volunteered resources reduce server overcommitment. Some redundant work remains because rule-based schedulers cannot anticipate availability and resource fluctuations. Overall, distribution reduces latency and dedicated capacity needs, showing the value of scheduler optimization.

\subsection{Limitations and Future Work}

This paper focuses on the collaborative distributed inference architecture and its modeling framework, instead of a production-ready implementation. Accordingly, we assume homogeneous users with identical availability and resource distributions, which also allows us to omit explicit fairness mechanisms. A real deployment could condition user behavior and resources on temporal and user-specific context and incorporate contribution-aware request prioritization. We further use \(1\,\mathrm{s}\) time steps from the server's perspective; irregular observations can be aligned with this representation through interpolation. 

Although the conceptual execution state grows quadratically with the user count, our implementation uses limited slots to bound the tasks maintained per node and scale linearly. Alternative request-centric representations could further reduce this space cost, making model scalability and decentralization promising research directions. The framework can also naturally extend to multiple dedicated servers, DAG-based workflows, and autoregressive workloads such as LLM inference. Finally, optimizing the scheduler and dedicated resource allocation formulated in Section~\ref{sec:problems}, as well as addressing deployment-level security and privacy, remains future work. These promising extensions build directly on the proposed model, whose modular factorization allows individual assumptions and components to be refined without redesigning the overall framework.

\section{Conclusion}

We proposed a collaborative distributed inference system that combines dedicated server infrastructure with resources contributed by users' devices, enabling demand to scale without proportional growth in dedicated capacity. To capture the system's stochastic and dynamic behavior, we developed a generative Markov model that supports simulation, analysis, and policy optimization. Our evaluation showed that distributed scheduling increasingly outperforms centralized inference as the user population grows, improving request completion and tail latency while substantially reducing dedicated resource consumption. These findings establish the feasibility of user-assisted inference and provide a foundation for future work on optimized scheduling, QoS-aware resource allocation, and extensions to richer workloads and deployment environments.

\begin{credits}
\subsubsection{\ackname} The computations were enabled by resources provided by the National Academic Infrastructure for Supercomputing in Sweden (NAISS), partially funded by the Swedish Research Council through grant agreement no. 2022-06725 and also supported by MSCA's TALENTS (101299722) project.
\end{credits}

\bibliographystyle{splncs04}
\bibliography{ref}

\end{document}